\documentclass[prd,twocolumn,reprint,preprintnumbers,nofootinbib]{revtex4-1}

\RequirePackage[colorlinks=true
,urlcolor=blue
,anchorcolor=blue
,citecolor=blue
,filecolor=blue
,linkcolor=blue
,menucolor=blue
,linktocpage=true
,pdfproducer=medialab
,pdfa=true
]{hyperref}

\usepackage{amsmath,amssymb,amsthm,amsfonts}
\usepackage{graphicx,tabularx}
\usepackage{float}
\usepackage{color}
\usepackage{multirow}  
\usepackage{cancel}
\usepackage{comment}

\usepackage{cleveref}
\crefname{section}{Sec.}{Secs.}
\crefname{figure}{Fig.}{Figs.}
\crefname{equation}{Eq.}{Eqs.}
\crefname{appendix}{Appendix}{Appendices}

\newcommand{\be}{\begin{equation} \begin{aligned}}
\newcommand{\ee}{\end{aligned} \end{equation}}

\newcommand{\bea}{\begin{eqnarray}}
\newcommand{\eea}{\end{eqnarray}}
\newcommand{\del}{\partial}

\begin{document}

\title{New Solutions for Rotating Boson Stars}
	
\preprint{{UCI-TR-2020-16}}
	
\author{Felix Kling}
\email{felixk@slac.stanford.edu}
\affiliation{SLAC National Accelerator Laboratory, 2575 Sand Hill Road, Menlo Park, CA 94025, USA}
	
\author{Arvind Rajaraman}
\email{arajaram@uci.edu}
\affiliation{Department of Physics and Astronomy, University of California, Irvine, CA 92697, USA}
	
\author{Freida Liz Rivera}
\email{flrivera@uci.edu}
\affiliation{Department of Physics and Astronomy, University of California, Irvine, CA 92697, USA}
	
\begin{abstract}
It has been shown that scalar fields can form gravitationally bound compact objects called boson stars. In this study, we analyze boson star configurations where the scalar fields contain a small amount of angular momentum and find two new classes of solutions. In the first case all particles are in the same slowly rotating state and in the second case the majority of particles are in the non-rotating ground state and a small number of particles are in an excited rotating state. In both cases, we solve the underlying  Gross-Pitaevskii-Poisson equations that describe the profile of these compact objects both numerically as well as analytically through series expansions.
\end{abstract}
	
\maketitle

\section{Introduction}
		
If light bosons, such as axions, form dark matter, it is potentially possible for them to collapse into bound compact objects, which are called boson stars~\cite{Kaup:1968zz, Ruffini:1969qy, Breit:1983nr} or axion stars~\cite{Barranco:2010ib, Braaten:2018nag, Eby:2019ntd}. Considerable work has been done in determining the wavefunctions for these compact objects, either numerically or semi-analytically in both non-relativistic and relativistic frameworks~\cite{Membrado:1989bqo, Moroz1998SphericallySS, Tod1999AnAA, Arbey:2003sj, Boehmer:2007um, Chavanis:2011zm, Chavanis:2011zi, Eby:2014fya, Eby:2015hsq, Mocz:2015sda, Kling:2017mif, Kling:2017hjm, Kan:2017uhj, Schiappacasse:2017ham, Eby:2017teq}. For a detailed comparison of the approximation methods and ansatz used in the literature see~\cite{Eby:2018dat}. 

Rotating boson star configurations  have also been studied, but all known solutions (that we have found in the literature) have the property that the total angular momentum increases proportionally to the mass of the star (e.g. \cite{Silveira:1995dh, Mielke:2016war, Davidson:2016uok, Jaramillo:2020rsv, Delgado:2020udb}).  In these solutions, the ratio of the angular momentum to the number of particles has a minimum value, and hence for a fixed number of particles, these solutions do not include configurations of  rotating boson stars with an arbitrarily small angular momentum.  

In this paper we remedy this gap, by finding new solutions which carry an arbitrarily small angular momentum for a fixed number of particles. We in fact find {\it two} different classes of such solutions.

Our first approach is a generalization of the solutions which exist in the literature, where all the particles are in the same state. However, we impose that the total angular momentum in the bosons is constrained to be fixed at a small value. This produces a state dominated by a spherical component, with a small admixture of a higher harmonic, naturally leading to a star with a small rotation. Our second approach is to take a small number of particles in the star to be in a higher spherical harmonic, while most of the particles are in the non-rotating state. Note that it is clear that such a solution must exist; for instance if a single particle is placed in a $\ell=1$ harmonic, there is no lower energy state with this angular momentum. We shall call these two ans\"atze respectively the one-state and two-state solution. We show that both these approaches successfully yield solutions for a rotating star with a small angular momentum. 

his paper is organised as follows: In order to set our notation and to connect to previous work, we first review our previous results for the case of non-rotating boson stars in \cref{sec:boson_stars}. We then turn to the rotating star: we consider the one-state ansatz in \cref{sec:one_state_solution} and the two-state ansatz in \cref{sec:two-state}. In each case, we set up the perturbation expansion around the non-rotating star, and solve the equations both numerically and in a series expansion, thereby providing strong numerical evidence that these solutions exist. We conclude in \cref{sec:conclusion}.

\section{Non-Relativistic Boson Stars}
\label{sec:boson_stars}

\subsection{Lagrangian and Structure Equations}
\label{sec:lagrangian}

Let us consider a real non-interacting scalar field $\phi(r,t)$   which is coupled to gravity. This scenario is described by the following Lagrangian 
\be
\label{eqn1}
    \!\! \mathcal{L}
    \!=\!\sqrt{g}\left[
    \frac{1}{16\pi G}R
    \!+\! \frac{1}{2}g^{\mu\nu} \partial_\mu\phi \partial_\nu\phi 
    \!-\! \frac{1}{2}m^2\phi^2 
\right] . 
\ee
The scalar field can form gravitational bound states, or boson stars. In this study, we focus on the case of dilute boson stars, which can be described by the Newtonian non-relativistic limit. For the case of QCD-axions, it has been shown that only dilute axion stars are stable over astronomical time scales~\cite{Visinelli:2017ooc,Chavanis:2017loo}. 

In the Newtonian limit, when the field $\phi$ couples only weakly to gravity, the metric can be written as $g_{\mu\nu} = \text{diag}(1+2\Phi,-1,-1,-1)$, where $\Phi$ is the Newtonian gravitational potential. We are interested in stationary solutions, in which case the gravitational potential is time independent. In this case the Ricci scalar takes the simple form $R =-2(\nabla \Phi )^2$. Also in the non-relativistic limit, we can treat the energy as being close to the mass, and we have $(\partial_t \phi)^2 \Phi = m^2 \phi^2{\Phi}$. The Lagrangian in \cref{eqn1} then becomes
\be
    \mathcal{L}
    &= \frac{1}{2} \frac{(\partial_t \phi)^2}{1+2\Phi} 
    - \frac{1}{2} (\nabla \phi)^2
    - \frac{1}{2} m^2 \phi^2
    - \frac{(\nabla \Phi)^2}{8 \pi G} \\
    &=  \frac{1}{2}(\del_\mu \phi\del^\mu\phi - m^2 \phi^2)
    - \frac{(\nabla \Phi)^2}{8 \pi G} 
    - m^2 \phi^2{\Phi}
    \label{eq:lag_nr}
\ee

Since the Lagrangian is quadratic in the scalar field, we can quantize the scalar in the usual way. We first find a set of wavefunctions satisfying 
\be
\nabla^2 \phi_n - 2m^2 \phi_n{\Phi} = -2mE_n\phi_n
\ee
and quantize by setting the scalar operator equal to
\be
\phi=\sum a_n^\dagger \phi_n +a_n \phi_n^* \ . 
\ee
The Hamiltonian is then
\be
H=\sum \sqrt{2mE_n+m^2} a_n^\dagger a_n 
\ee
The eigenstates are of the form 
\be
\Psi = a_{n_1}^\dagger a_{n_2}^\dagger ..|0\rangle \ . 
\ee

The gravitational potential interacts with the scalar through the term $m^2 \phi^2{\Phi}$. This leads to the equation for the potential
\be
    \nabla^2 \Phi = 4 \pi G m \, \langle \Psi  |\phi^2|\Psi\rangle 
\ee
These field equations, often referred to as Gross-Pitaevskii-Poisson equations, are the structure equations for the boson star. 

\subsection{The non-rotating Boson Star}
\label{sec:dimensionless}

For the non-rotating star, we consider an ansatz where we have $N$  particles in the ground state $\psi_{nr}\equiv \phi_0$, which has an energy eigenvalue $e_{nr}\equiv E_0$. The state is then
\be
    \Psi_{nr}= \frac{1}{N!} (a_{0}^\dagger)^N |0\rangle
\ee
and the corresponding structure equations are given by a Schr\"odinger type equation for the ground state wavefunction 
\be
    \nabla^2\psi_{nr}-2m^2 \Phi_{nr} \psi_{nr}=-2me_{nr}\psi_{nr}
    \label{eq:schrodinger}
\ee
and a Poisson equation for the gravitational potential
\be
    \nabla^2 \Phi_{nr} =  4 \pi G N m \, |\psi_{nr}|^2 \ . 
    \label{eq:newton}
\ee

To solve the structure equations for the boson stars, it is convenient to introduce dimensionless variables. Following  Refs.~\cite{Kling:2017mif, Kling:2017hjm}, we define 
\be
    s_0  &= - \frac {\sqrt{2\pi GM}}{e_{nr}}\psi_{nr} \ , 
    &v_0 &= -1 + \frac{m}{e_{nr}} \Phi_{nr} \ , \\ 
    z & = \sqrt{-2me_{nr}} \ r \ , 
    &e_{nr}&=- \frac{G^2 M^2 m^3}{2\beta^2} . 
    \label{eq:dimless_vars}
\ee
where $M=Nm$ is the star's mass. Using these variables, we can rewrite the Gross-Pitaevskii-Poisson equations as
\vspace{-4mm}
\be
    \nabla_z^2 s_0 = - s_0v_0
    \quad \text{and} \quad 
    \nabla_z^2 v_0 = - |s_0|^2
    \label{eq:structure_nr}
\ee
where the derivatives $\nabla_z$ are now with respect to the dimensionless coordinate $z$. In dimensionless variables, we can write the normalization condition of the wavefunction, $\int |\psi|^2 dV=1$, as $\int s_0^2 z^2 dz = 2 \beta$ and associate $2\beta$ with the mass of the star. Note that up to scalings, there is only one ground state solution for non-interacting boson stars.
\medskip

In \cite{Kling:2017mif, Kling:2017hjm} we have solved the Gross-Pitaevskii-Poisson equations and obtained a semi-analytic solution for the ground state of the boson star. In this approach,  the profiles at both small and large radii are separately described through a series expansion of the wavefunction and potential and matched at an intermediate point. At small radii, the profile can be described by an even polynomial around the center of the star ($z=0$)
\be
    s_0^{\text{near}} = \sum_{n=0}^{\infty} s^0_n z^n \quad \text{and} \quad
    v_0^{\text{near}} = \sum_{n=0}^{\infty} v^0_n z^n \ .
\label{smallz-expansion}
\ee
At large radii, we take 
\be
    s_0^{\text{far}} &= \sum_{n,m=0,0}^{\infty,\infty}  s^0_{n,m} 
    \left( \frac{e^{-z}}{z^\sigma} \right)^n   z^{-m} 
    \quad \text{and} \quad \\
    v_0^{\text{far}} &=  \sum_{n,m=0,0}^{\infty,\infty} v^0_{n,m}
    \left( \frac{e^{-z}}{z^\sigma} \right)^n  z^{-m} \ .
\label{largez-expansion}
\ee

The potential and wavefunction are fully specified by knowing the parameters of the leading expansion
\be
    &s_0^{\text{near}} \approx s^0_0 + \dots, \quad
    s_0^{\text{far}}    \approx \alpha e^{-z}z^{\beta-1} + \dots \\
    &v_0^{\text{near}} \approx v^0_0 + \dots, \quad 
    v_0^{\text{far}}    \approx -1 + \frac{2\beta}{z}  + \dots \ .
\ee
The remaining coefficients can be obtained using recursion relations which can be derived from the Gross-Pitaevskii-Poisson equations and have been presented in \cite{Kling:2017mif, Kling:2017hjm}. We can determine the four expansion parameters either through a fit to the numerical solution, or by matching the small and large radius wavefunction and their derivatives at a matching point $z^*$. We have obtained the following solutions \cite{Kling:2017mif} 
\be
    s^0_0     &=  1.02149303631 \pm 1.4 \cdot 10^{-10} \\
    v^0_0     &=  0.93832284019 \pm 1.3 \cdot 10^{-10} \\
    \alpha  &=  3.4951309897\phantom{9}  \pm 5.1 \cdot 10^{-9} \\
    \beta   &=  1.7526648513\phantom{9}  \pm 1.3 \cdot 10^{-9} \ . 
\label{numerics-paramaters}
\ee

\subsection{Slowly Rotating Boson Stars}

We now turn to a study of rotating boson stars. In particular, we look for slowly rotating boson stars solution, which can be treated as a perturbation around the non-rotating solution. That is, the non-rotating solution should admit a normalizable perturbation such that the perturbation carries angular momentum. The existence of such a perturbation would indicate that a slowly rotating boson star can be found at least at the linearized level, which is suggestive that the full solution should exist.

To look for these states, we impose a constraint on the total angular momentum of the boson star 
\be
	N \int \phi \, \hat{L}^2 \, \phi \, dV 
	= L_{\text{star}}^2\neq 0
	\label{eq:constraint}
\ee
where $\hat{L}^2$ is the usual total angular momentum operator $\hat{L}^2=  \partial_\theta^2 + (\sin\theta)^{-2} \partial_\phi^2$. On astrophysically relevant times scales, the boson star's angular momentum $L_\text{star}$ is a fixed quantity. We implement this constraint by introducing a Lagrange multiplier $\mu$. The Lagrangian in \cref{eq:lag_nr} then becomes
\be
    \mathcal{L} &=
    \frac{1}{2}( \partial_\mu \phi \partial^\mu \phi -  m^2 \phi^2)
    - \frac{(\nabla \Phi)^2}{8 \pi G} \\
    & - m^2 \phi^2{\Phi} 
    + m \mu \left(N \phi \hat{L}^2\phi 
    - L_{\text{star}}^2 \phi^2 \right) \ . 
\ee

We can repeat the quantization procedure and find
resulting equations of motion are the Poisson equation given in \cref{eq:newton} and a modified Schr\"odinger-type equation
\be
    \frac{1}{2m} \nabla^2 \phi_n & = 
	(m \Phi \!-\! E_n)  \phi_n
	- \mu \hat{L}^2 \phi_n 
	+ \frac{\mu L^2_{\text{star}}}{N}  
	\phi_n . 
    \label{eq:schrodinger1}
\ee
In the following, we will present two possible solutions for the slowly rotation boson star, and obtain the corresponding ground state wave-function.

\section{Rotating Boson Stars: One-state solution}
\label{sec:one_state_solution}
	
\subsection{The Ansatz }
\label{sec:oss_ansatz}

We first look for a solution where all the particles are in the ground state. The state is then
\be
    \Psi= \frac{1}{N!}(a_{0}^\dagger)^N |0\rangle
\ee
This is formally similar to the non-rotating case, but because of the constraints, we must take the ground state in this sector to have non-zero angular momentum.

We take the ground state to be a perturbation around the non-rotating spherically symmetric solution $\psi_{nr}(r)$ obtained in \cref{sec:dimensionless}. In particular, we choose an ansatz in which the wavefunction and potential perturbation are expanded in spherical harmonics $Y_{\ell m}$ with $\ell\geq1$ and $m=0$,
\be
	\phi_0(r,\theta,\phi) &=\psi_{nr}(r)+\epsilon\psi_1(r)Y_{\ell0}(\theta)\ , \\
	\Phi(r,\theta,\phi) &=\Phi_{nr}(r)+\epsilon\Phi_1(r)Y_{\ell0}(\theta)\ , 
\ee
as well as $-E_0=e_{nr} + \epsilon e_1$. The expansion parameter $\epsilon$ is taken to be parametrically small, which allows us to work in linear order perturbation theory. 

The angular momentum constraint in \cref{eq:constraint} relates the value of $\epsilon$ and the star's angular momentum, such that $\epsilon =  L_{\text{star}} \times [N \ell (\ell+1) \cdot \int |\psi_1|^2 dV ]^{-1/2}$. We then find that the last term in \cref{eq:schrodinger1} is of order $L^2_{\text{star}} \sim \epsilon^2$ and can therefore be ignored at linear order in perturbation theory.

We now insert this ansatz into the field equations \cref{eq:schrodinger1} and \cref{eq:newton}. Collecting terms at zeroth order in $\epsilon$, we recover the equations of motion for a non-rotating boson star, whose solution we presented in \cref{sec:dimensionless}. Matching the terms proportional to $\epsilon Y_{\ell0}$ we find the structure equations for the perturbation 
\be
	\!\!\frac{1}{2m}\nabla^2 \psi_1
	\!&=\! m \Phi_1 \psi_{nr} 
	\!+\! (m \Phi_{nr} \!-\! e_{nr}) \psi_1  
	\!-\! \mu \ell (\ell\!+\!1) \psi_1  
	\\
	\nabla^2 \Phi_1 \!&=\!  4 \pi G N m (\psi^*_{nr} \psi_1 
	\!+\! \psi^*_1 \psi_{nr})  \ . 
	~~~~~~~~~~~~
    \label{eq:eom1_phi}
\ee

Finally, collecting the terms proportional to $\epsilon Y_{00}$ implies $e_1=0$, meaning that the rotation does not induce a shift in the binding energy at leading order in perturbation theory. Such a shift first appears at order $\epsilon^2$. 

We perform the change of variables in \cref{eq:dimless_vars} and further  define
\be
    \!\!\!
    s_{1} \!=\! \left[\frac{2\pi GM}{e_{nr}^2} \right]^{\frac{1}{2}}
    \!\!\! \psi_{1}, 
    \ \ \ \ 
    v_1 \!=\! \frac{m}{e_{nr}}\Phi_1 , 
    \ \ \ \ 
    \Gamma \!=\! \frac{\mu\ell(\ell\!+\!1)}{e_{nr}} .
\label{eq:re-scale2}
\ee
The resulting structure equations for the dimensionless field and potential perturbations $s_1$ and $v_1$ then read
\be
	\nabla_z^2 s_1 \!-\! \ell( \ell\!+\!1) /z^2 \, s_1
	&= - v_1  s_0  
	\!-\! v_0  s_1 
	\!+\! \Gamma s_1 
	\\
	\nabla^2_z v_1  \!-\! \ell (\ell\!+\!1) /z^2 \, v_1
	&=   - 2 s_0 s_1 \ . 
\label{eq:structure_oss}
\ee

\subsection{Series Expansion }
\label{sec:oss_series}
 
We have seen in \cref{sec:dimensionless} that we can describe the profile of the non-rotating boson star through an infinite series for the wavefunction and potential. We will follow the same approach to obtain a solution for \cref{eq:structure_oss}. 

At small radii, the profiles for $s_1$ and $v_1$ can be described via a polynomial around the center of the boson star $z=0$,
\be
    s_1^\text{near} = \sum_{n=0}^{\infty} s_n^1 z^n \;\;\;\;\text{and}\;\;\;\;
    v_1^\text{near} = \sum_{n=0}^{\infty} v_n^1 z^n\ .
\label{smallz-expansion_oss}
\ee
By matching the coefficients in \cref{eq:structure_oss} we obtain the recursion relations
\be
    \left[ (n\!+\!2)(n\!+\!3) \!-\! \ell(\ell\!+\!1) \right] s^1_{n+2} 
    & = \Gamma s_n^1 \\ 
    - \sum_{m=0}^{n} [s^0_m v^1_{n-m} + & s^1_m v^0_{n-m}] \\
    \left[ (n\!+\!2)(n\!+\!3) \!-\! \ell(\ell\!+\!1) \right] v^1_{n+2} 
    &= \!-\!\sum_{m=0}^{n} s^0_m s^1_{n-m} 
\ee
Requiring the left hand side of \cref{eq:structure_oss} to be defined at $z=0$ implies that the perturbation vanishes at the origin and hence $s^1_0 = v^1_0 = 0$. The profile at small radii can therefore be fully parameterized in terms of the derivative of the wavefunction and potential at the origin $\partial_z s_1 = s_1^1 $ and $\partial_z v_1 = v_1^1$. 

At large radii, we will once again use the series expansion ansatz
\be
  	s^{\text{far}}_{1} &=\sum\limits_{n,m=0,0}^{\infty,\infty} {s}_{n,m}^{1} 
  	\left (\frac{e^{-z}}{z^{\sigma }}  \right )^{n} z^{-m}  
  	\ \ \ \text{and} \\
   v^{\text{far}}_{1} &=\sum\limits_{n,m=0,0}^{\infty,\infty} {v}_{n,m}^{1} 
   	\left (\frac{e^{-z}}{z^{\sigma }}  \right )^{n} z^{-m} \ . 
   	\label{eq:rot_farexp}
\ee
and obtain the recursion relations 
\be
    \label{eq:oss_recs}
    & n^{2} {s}_{n,m}^{1} 
    + 2n (n\sigma \!+\! m \!-\! 2) {s}_{n,m-1}^{1} \\
    &+ [(\sigma n \!+\! m \!-\!2)(\sigma n \!+\! m \!-\!3) 
    - \ell(\ell\!+\!1)]{s}_{n,m-2}^{1} \\
    &= \!-\!\!\!\!\!\sum_{p,q=0,0}^{n,m} \!\!\! s^0_{p,q} v^1_{n-p,m-q} 
    - \!\!\!\!\!\sum_{p,q=0,0}^{n,m} \!\!\! s^1_{p,q} v^0_{n-p,m-q} 
    + \Gamma s^1_{n,m} \\
\ee 
and
\be
    & n^{2} v^1_{n,m}  
    + 2n (n\sigma \!+\! m \!-\! 2) {v}_{n,m-1}^{1} \\
    &+ [(n\sigma\!+\!m\!-\!2)(n\sigma\!+\!m\!-\!3)
    - \ell(\ell\!+\!1)] v^1_{n,m-2} \\
    & = - 2 \!\!\!\sum_{p,q=0,0}^{n,m} \!\!\! s^0_{p,q} s^1_{n-p,m-q} \ . 
    \label{eq:oss_recv}
\ee

Let us note the following properties of $s_1^\text{far}$ and $v_1^\text{far}$:
i) \cref{eq:oss_recv} with $n=0$ implies that $v^1_{0,m}=0$ unless $m = \ell+1$. At large radius, the potential is then approximately described by $v^\text{far}_1 = v^1_{0,\ell+1} z^{-(\ell+1)}$, while all other terms in the expansion are at least exponentially suppressed.
ii) Normalizability of the wavefunction requires $s^1_{0,0}=0$. \cref{eq:oss_recs} with $n\!=\!0$ then implies that all coefficients $s^1_{0,m}$ vanish as  well. This means that similar to the non-rotating wavefunction $s_0^\text{far}$, the wavefunction of the rotating perturbation $s_1^\text{far}$ decays at least exponentially. 
iii) \cref{eq:oss_recs} and \cref{eq:oss_recv} further imply that the potential contains only non-vanishing components $v^1_{n,m}$ for even $n$ while the wavefunction only has non-vanishing component $s^1_{n,m}$ for odd $n$.

The first non-vanishing terms for wavefunction $s_1^\text{far}$ appear for $n\!=\!1$. Using the known $n\!=\!0$  solutions of the non-rotating case, we can simplify \cref{eq:oss_recs} and write 
\be
	2&(\beta \!+\! \sigma\!+\!m\!-\!2) s^1_{1,m\!-\!1} 
	+ s^0_{1,m-\ell-1} v^1_{0,\ell+1} 
	- \Gamma s^1_{1,m} \\
	& = -[(\sigma\!+\!m\!-\!2)(\sigma\!+\!m\!-\!3)-\ell(\ell\!+\!1)] s^1_{1,m\!-\!2} 
	\label{eq:oss_recs_n1} 
\ee

Setting $m\!=\!0$, \cref{eq:oss_recs_n1} can be written as $s^1_{1,0}\,\Gamma =0$, which either implies $s^1_{1,0}=0$ or $\Gamma=0$. Although both possibilities will lead to a solution, we will mainly focus on the $\Gamma\!=\!0$ solution. For $m\!=\!1$, we find that $1-\sigma = \beta$, where the $\sigma$ originates from the $s_1^\text{far}$ expansion. This is the same relation we found for the non-rotating boson star in Ref.~\cite{Kling:2017mif}, justifying our ansatz to use the same $\sigma$ for both the $s_0^\text{far}$ expansion in \cref{largez-expansion} and $s_1^\text{far}$ expansion in \cref{eq:rot_farexp}. Finally, setting $m\!=\!M+1$ we can use \cref{eq:oss_recs_n1} to obtain the recursion relation
\be
    \!\!2 M s^1_{1,M}  
	&= [\ell(\ell\!+\!1) 
	\!-\! (\sigma\!+\!M\!-\!1)(\sigma\!+\!M\!-\!2)] s^1_{1,M-1}  \\
	&- v^1_{0,\ell+1} \ s^0_{1,M-\ell} \ . 
\ee
This means, that all coefficients can be determined recursively from $s^1_{1,0}$ and $v^{1}_{0,\ell}$. More generally, we can use \cref{eq:oss_recs} and \cref{eq:oss_recv} to recursively calculate all coefficients $s^1_{n,m}$ and $v^1_{n,m}$ in the expansion of $s^\text{far}_1$ and $v^\text{far}_1$.

We can now determine the expansion parameters by matching the near and far field wavefunction and potential and their derivatives at a matching point $z^*$. We have performed such a matching using the near field solution in \cref{smallz-expansion_oss} truncated with $n \leq 100$ and the far field solution in \cref{eq:rot_farexp} truncated with $n \leq 5$, $m \leq 5$ and obtained
\be
    s^1_1     &= 0.91848 \pm 0.00061 \ , \\
    s^1_{1,0} &= 10.125 \pm 0.052 \ , \\
    v^1_{0,2} &= 10.111 \pm 0.056 \ . 
\label{eq:oss_matching}
\ee
where the perturbation is normalized such that $v^1_1 = 1$. To estimate the uncertainty associated with the matching procedure, we performed multiple matchings for $3 \leq z^* \leq 3.5$. 

\subsection{Leading Order Analytic Far-Field Solution}
\label{sec:oss_whittaker}

\begin{figure*}[t]
\centering
    \includegraphics[trim=2 20 50 40,clip,width=0.49\textwidth]
    {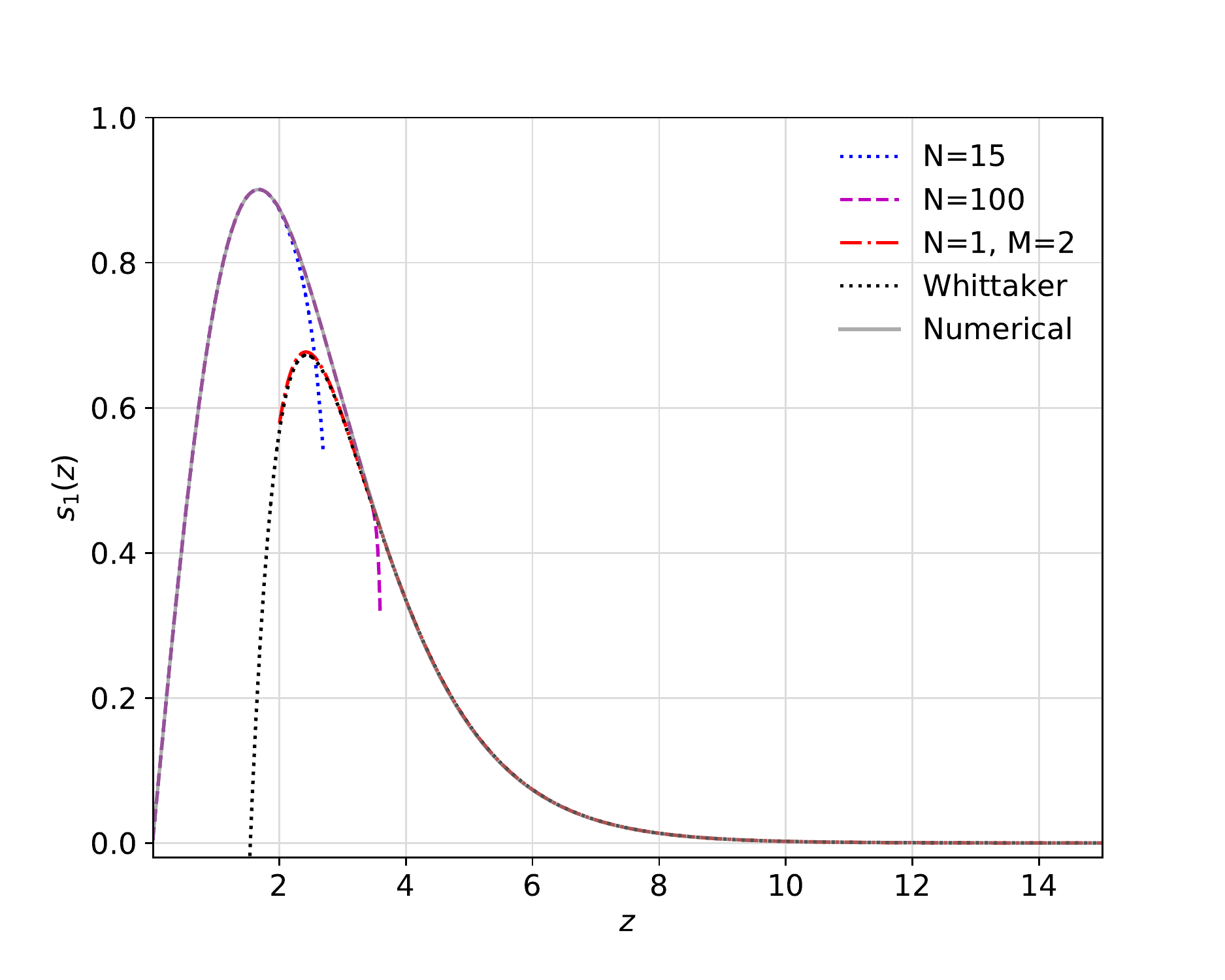}
    \includegraphics[trim=2 20 50 40,clip,width=0.49\textwidth]
    {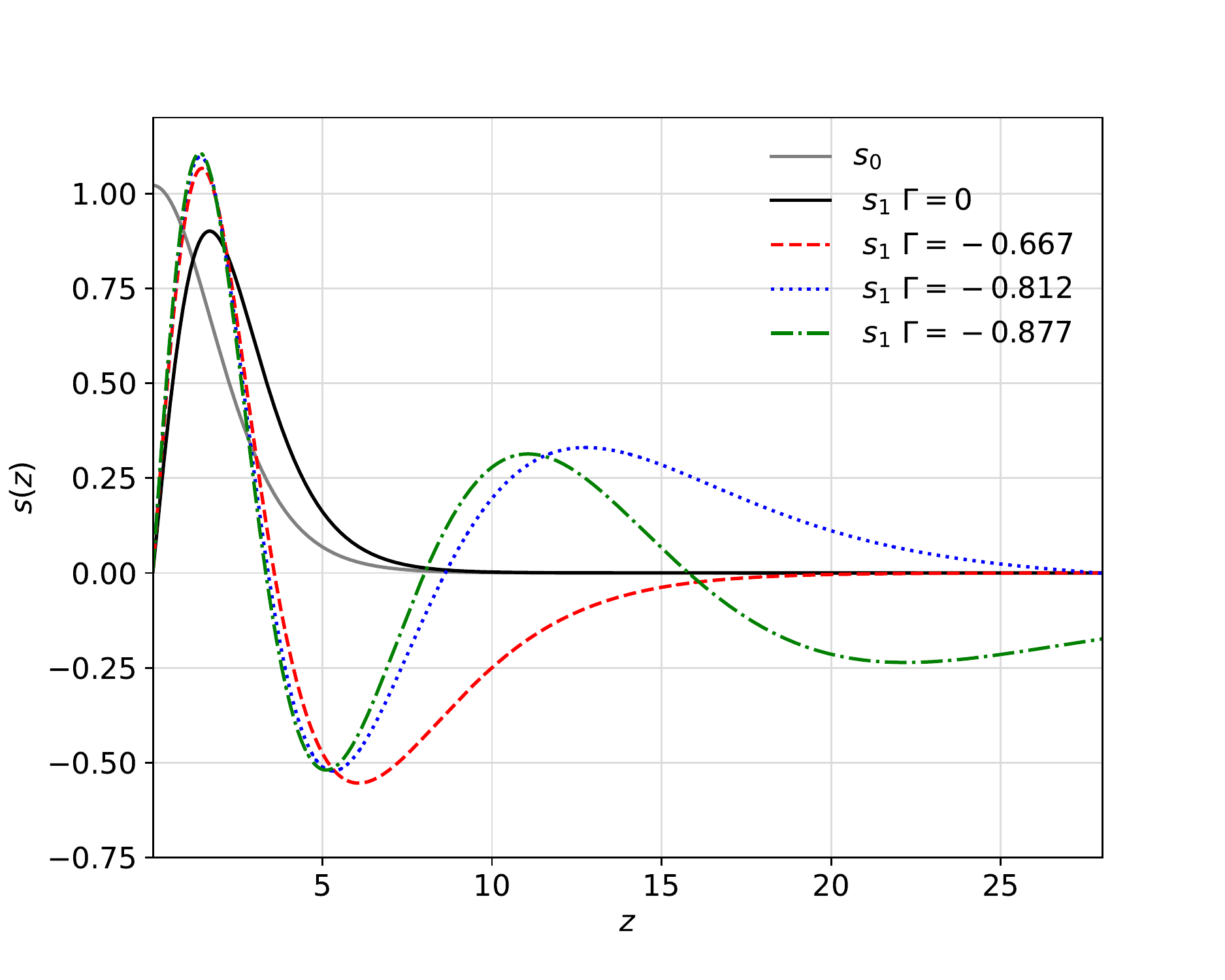}
\caption{One-state solution. \textbf{Left:} The numerical solution (solid gray) and truncated series expansion at small radius (dotted blue and dashed magenta) and large radii (dot-dashed red) as well as the analytic Whittaker approximation (dotted black) for the wavefunction of a rotating perturbation with $\ell=1$ and $\Gamma=0$. 
\textbf{Right:} Numerical solution for the wavefunction of a rotating perturbation with $\ell=1$ and $\Gamma=0$ (black), $\Gamma  =-0.667$ (dashed red), $\Gamma=-0.812$ (dotted blue) and  $\Gamma=-0.877$ (dot-dashed green).  The non-rotating ground state wavefunction is shown for comparisons (gray solid).
}
\label{fig:one_state}
\end{figure*}

Similar to the ground-state, it is also possible to obtain an approximate analytical solution for the far field at leading order $n=1$. Using that $v_0 \approx -1 + 2\beta z^{-1}$ and $v_1 \approx v^1_{0,\ell+1} z^{-\ell+1}$, we can write the Schr{\"o}dinger-like equation in \cref{eq:structure_oss} as
\be
    \nabla^2_z s_1 
    \- \frac{ \ell (\ell\!+\!1) }{z^2} s_1 
    = \left (1 + \Gamma - \frac{2\beta}{z}\right) s_1 - \frac{v^1_{0,\ell+1}}{z^{\ell+1}} s_0\ . 
\ee
After performing a change of variables to $w=2 z s_1$, $w_0 = 2 z s_0$ and $y=2z (1+\Gamma)^{1/2}$ we can write
\be
    \frac{d w^2}{dy^2} 
    + \left(-\frac{1}{4}+\frac{\kappa}{y} - \frac{\ell(\ell\!+\!1)}{y^2}\right) w 
    = - \frac{\lambda}{y^{\ell-1}} \frac{v^1_{0,\ell+1}}{y^2} w_0
    \label{eq:inhom_whittaker}
\ee
with $\kappa = \beta \cdot (1\!+\!\Gamma)^{-1/2}$ and $\lambda = [2(1\!+\!\Gamma)^{1/2}]^{\ell-1} $. Looking at the homogeneous part on the left hand side, we rediscover the \textit{Whittaker equation}. Following the notation of Ref.~\cite{Abramowitz:1974}, the solution to \cref{eq:inhom_whittaker} is given by a linear combination of the \textit{Whittaker functions} $W_{\kappa,\mu}(y)$ and $M_{\kappa,\mu}(y)$ as well as one solution $w^{in}_{\kappa,\mu}(y)$ to the inhomegeneous Whittaker equation 
\be
 w(y)=c \cdot W_{\kappa,\mu}(y)+c' \cdot M_{\kappa,\mu}(y) + w^{in}_{\kappa,\mu}(y)
\ee
where $\mu^2\!=\!\frac{1}{4} \!+\! \ell(\ell\!+\!1)$ or $\mu \!=\! \ell\!+\!\frac{1}{2}$. 

For $\Gamma\!=\!0$, and hence $\kappa\!=\!\beta$, normalizability of the wavefunction requires $c'\!=\!0$. For $\ell=1$, we also see that $w = v^1_{02} [\ell (\ell+1)]^{-1} w_0$ is a solution of the innomogeneous Whittaker equation in \cref{eq:inhom_whittaker}. This then implies that 
\be
    s^{\text{far},\ell=1}_1 
    = \frac{c}{2z} W_{\beta,\mu}(2z) 
    + \frac{v^1_{0,2}} {2} \frac{\alpha}{2^{\beta} z}  W_{\beta,\frac{1}{2}}(2z) 
\label{eq:oss_whittaker}
\ee
Expanding the Whittaker function, we obtain $s^\text{far}_1 = \left(c 2^{\beta-1} + \alpha v^1_{02}/2 \right) e^{-z} z^{-\sigma}+ \dots$, which allows us to identify $c= \left(2 s^1_{1,0} - \alpha v^1_{02} \right) 2^{-\beta}$. 

For $\Gamma \neq0$, additional normalizable solutions with $c'\neq 0$ could exist. The Whittaker function $M_{\kappa,\mu}(y)$ converges to zero for large values of $y$ if $\kappa$ is a natural number $\geq2$, fixing the corresponding values of $\Gamma = \beta^2 / \kappa^2 - 1$.

\subsection{Numerical analysis}

In \cref{sec:oss_series} we have shown that the wavefunction and potential profile of the rotating boson star can be described by a series expansion, which is characterized by the expansion parameters given in \cref{eq:oss_matching}. In the following, we will compare this result to the numerical solution of \cref{eq:structure_oss}, focusing on the case $\ell=1$. 

As we have seen before, near $z=0$ the solution takes the form $s_1\sim s^1_1 z+\dots$ and $v_1\sim v^1_1 z+\dots$. To obtain a numerical solution, it is convenient to normalize the field $s_1$ and the potential $v_1$ such that $v^1_1 = 1$, so that the solution is only parameterized by $s^1_{1}$. Using a Runge-Kutta~4 method, we then perform the numerical integration of \cref{eq:structure_oss}. For most values of $s^1_1$, the wavefunction profile will diverge to positive or negative infinity at large radii $z\gg 1$.  Using a shooting point method analogous to those used by the authors of Ref.~\cite{Tod1999} and \cite{Kling:2017mif}, we adjust $s^1_1$ such that the wavefunction converges and becomes square integrable.

The numerical solution for $\Gamma=0$ is shown in the left panel of \cref{fig:one_state} as solid gray line. Fitting the solution by the far potential $v^\text{far}_1 \approx v^1_{0,2} z^{-2}$ and the far wavefunction given in \cref{eq:oss_whittaker}, we can extract the expansion parameters of the series expansion
\be
    s^1_1    &= 0.91835 \pm 0.00014 \ , \\
    s^1_{1,0} &= 10.123 \pm 0.018 \ , \\
    v^1_{0,2} &= 10.080089 \pm 0.000035 \ . 
\ee
where the uncertainties were obtained by varying the fit range. These results agree with our previous findings based on the matching between the near and far solution obtained in \cref{eq:oss_matching}. 

The dashed curves show the wavefunction profile of the truncated near solution in \cref{smallz-expansion_oss} with $n \leq 15$ and $n \leq 100$ as well as the far solution of \cref{eq:rot_farexp} with $n\leq 1$ and  $m\leq 2$. Here the truncated solution takes the simple form 
\be
    \hspace{-0.3cm}
    \!\!s_1 = 
    \begin{cases}
        0.918 z  - 0.188 z^{3} + 0.024 z^{5} & \\
        \ - 0.262 \!\cdot\! 10^{-2} z^{7} \,\; + 0.251 \!\cdot\! 10^{-3} z^{9} 
        & \ \ \ \text{for}\\
        \ - 2.239 \!\cdot\! 10^{-5} z^{11} + 1.896 \!\cdot\! 10^{-6} z^{13} 
        & z\!<\!3.45 \!\!\! \\
        \ - 1.547 \!\cdot\! 10^{-7} z^{15} & \\
        &\\
        10.12 \, z^{0.752} e^{-z} - 14.16 \, z^{-0.247}e^{- z} 
        & \ \ \ \text{for}\\
        \ - 1.93 \, z^{-1.247}  e^{- z} 
        &  z\!>\!3.45 \!\!\! \\
    \end{cases}
\ee
We can see that already such few terms in the series expansion are sufficient to describe the wavefunction well. The dotted black curve shows the Whittaker function solution of \cref{eq:oss_whittaker}, which is already well described by the first few terms of the far field expansion. 

The right panel of \cref{fig:one_state} shows the numerical solution for both $\Gamma = 0$ and $\Gamma \neq 0$, alongside with the non-rotating ground-state solution $s_0$ discussed in \cref{sec:dimensionless}. In particular, we found that solutions exist for $\Gamma = -0.667, -0.812, -0.887$. These values are consistent with the relation $\Gamma = \beta^2 / \kappa^2 - 1$ found in \cref{sec:oss_whittaker} for $\kappa=3,4,5$. Notably, $\kappa-2$ also characterizes at how many radii the wavefunction vanishes identically, $s_1 = 0$. 
	
\section{Rotating Axion Stars: two-state solutions}
\label{sec:two-state}
	
\subsection{The Ansatz}
	
We now consider a second approach to find rotating boson star solutions. In this ansatz, we look for a state where $N$ particles are in the ground state $a_{0}^\dagger|0\rangle$, and $k$ particles are in the excited state $a_{1}^\dagger|0\rangle$. The state is then
\be
    \Psi= \frac{1}{N!k!}(a_{0}^\dagger)^N 
    (a_{1}^\dagger)^k |0\rangle
\ee
This leads to the Poisson-type equation for the potential
\be
    \nabla^2 \Phi = 4 \pi G m ( N|\phi_0|^2+k|\phi_1|^2 )
\ee
which should be solved along with the Schr\"odinger-type equations
\be
    \nabla^2 \phi_0 - 2m^2 \phi_0{\Phi}	
    &=-2mE_0\phi_0 \ , 
    \\
    \nabla^2 \phi_1 - 2m^2 \phi_1{\Phi} + 2m\mu \hat{L}^2 \phi_1
	&=-2mE_1\phi_1 \ .
	\label{eq:tss_schrodinger}
\ee
We will assume $k\ll N$, and perturb in the small parameter $\epsilon=k/N$. For this reason, we dropped the $L_\text{star}^2$ term in \cref{eq:tss_schrodinger} which only contributes at subleading order in $\epsilon$. Now, to zeroth order in  $\epsilon$, $\Phi$ will just be equal to the potential for the non-rotating star $\Phi_{nr}$, $\phi_0$ is equal to the wavefunction for the non-rotating star $\psi_{nr}$, and $E_0=e_{nr}$. As before, we will consider a single $Y_{\ell 0}$ mode i.e. 
\be
    \phi_1(r,\theta,\phi)= \psi_1(r) Y_{\ell 0}(\theta,\phi)
\ee

We again perform the change of variables in \cref{eq:dimless_vars} and further define 
\be
    s_{1} = \left[\frac{2\pi GM}{e_{nr}^2} \right ]^{\frac{1}{2}} \psi_{1},
    \quad \quad
    \Gamma=\frac{E_1 \!+\! \mu \ell(\ell\!+\!1)}{e_{nr}}-1
\ee
and obtain the structure equation
\be
	\nabla_z^2 s_1 \!-\! \ell( \ell\!+\!1) /z^2 \, s_1=  - v_0  s_1 
	+\Gamma s_1
	\label{eq:structure_oss2}
\ee 

The angular momentum of the boson star is equal to $L_\text{star} = [k \cdot  \ell (\ell+1) \cdot \int |\psi_1|^2 dV ]^{1/2}$. Note that unlike for the one-state case, in this case the rotation does induce a shift in the binding energy at leading order in perturbation theory.

\subsection{Series Expansion }

As before, we will parameterize the wavefunction via an infinite series expansion. At small radii, the profile for $s_1$ can be described via a polynomial around the center of the boson star $z=0$,
\be
    s_1^\text{near} = \sum_{n=0}^{\infty} s_n^1 z^n \ . 
    \label{smallz-expansion_oss2}
\ee
By matching the coefficients in \cref{eq:structure_oss2} we obtain the recursion relation
\be
    \!\!\left[ (n\!+\!2)(n\!+\!3) \!-\! \ell(\ell\!+\!1) \right] s^1_{n+2} 
    = \Gamma s_n^1 
    \!-\!\! \sum_{m=0}^{n} \! s^1_m v^0_{n-m} 
\ee
As in the one-state case, requiring the left hand side of \cref{eq:structure_oss2} to be defined at $z=0$ implies that the perturbation vanishes at the origin and hence $s^1_0 = 0$. The profile at small radii can therefore be fully parameterized in terms of the derivative of the wavefunction at the origin $\partial_z s_1 = s_1^1 $. 

At large radii, we will use the series expansion
\be
  	\!s^{\text{far}}_{1}
  	\!&=\!\!\!\!
  	\sum\limits_{n,m=0,0}^{\infty,\infty} {s}_{n,m}^{1} 
  	\left(\frac{e^{-\sqrt{1+\Gamma}z}}{(\sqrt{1+\Gamma}z)^{\sigma' }} 
  	\right)^{n} 
  	\!\!\!(\sqrt{1+\Gamma}z)^{-m} . \!\!
   	\label{eq:rot_farexp2}
\ee
Note that the form of this ansatz is slightly different than for the non-rotating boson star in \cref{largez-expansion} and the one-state solution in \cref{eq:rot_farexp}. As we will see later, two-state solutions only exist for $\Gamma \neq 0$, and the resulting far field solution would approximately follow the Whittaker function $W_{\kappa,\mu}(2z \sqrt{1+\Gamma})$. In order to match the asymptotic behaviour of this Whittaker function solution, the additional $\sqrt{1+\Gamma}$ factor as well as a new parameter $\sigma'$ have been included in the series expansion ansatz.

The coefficients of the expansion are related by the recursion relation
\be
    \label{eq:oss_recs2}
    &(1+\Gamma) 
    \big(n^{2} {s}_{n,m}^{1} 
    + 2 n (n\sigma' \!+\! m-2) {s}_{n,m-1}^{1}  \\
    &+ [ (\sigma' n \!+\! m \!-\! 2) 
    (\sigma' n \!+\! m \!-\! 3 ) 
    - \ell(\ell \!+\! 1)]{s}_{n,m-2}^{1}\big)  \\
    &= - {s}_{n,m}^{1} {v}_{0,0}^{0} 
    - \sqrt{1+\Gamma} {s}_{n,m-1}^{1} {v}_{0,1}^{0}  
    + \Gamma  {s}_{n,m}^{1} \ . 
\ee
Here we have used the approximate ground-state potential $v^{far}_0 \approx -1 + 2\beta / z$, such that the Cauchy product $v_0 s_1$ is well defined. 
\medskip

\begin{figure*}[t]
\centering
    \includegraphics[trim=2 20 50 40,clip,width=0.49\textwidth]
    {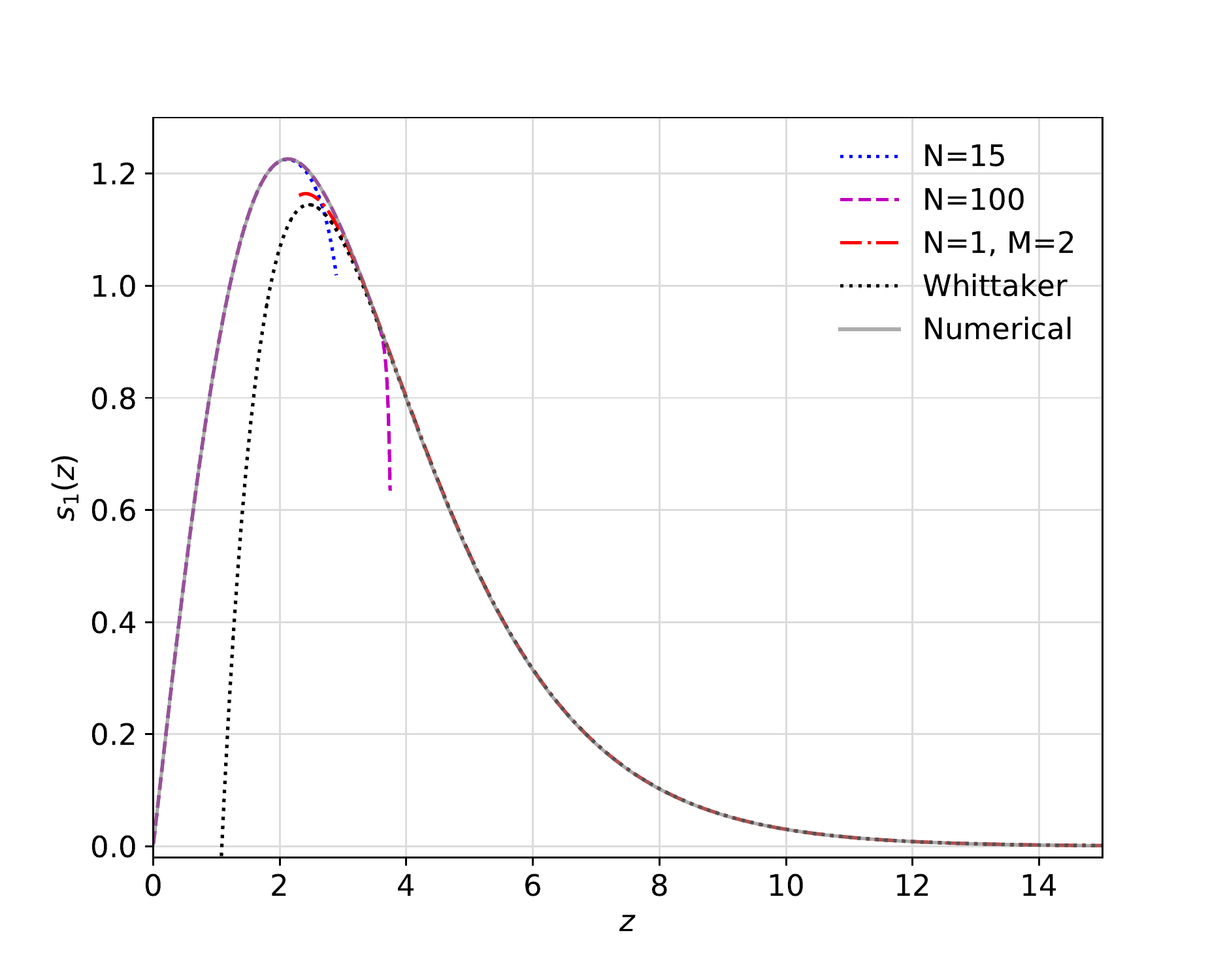}
    \includegraphics[trim=2 20 50 40,clip,width=0.49\textwidth]
    {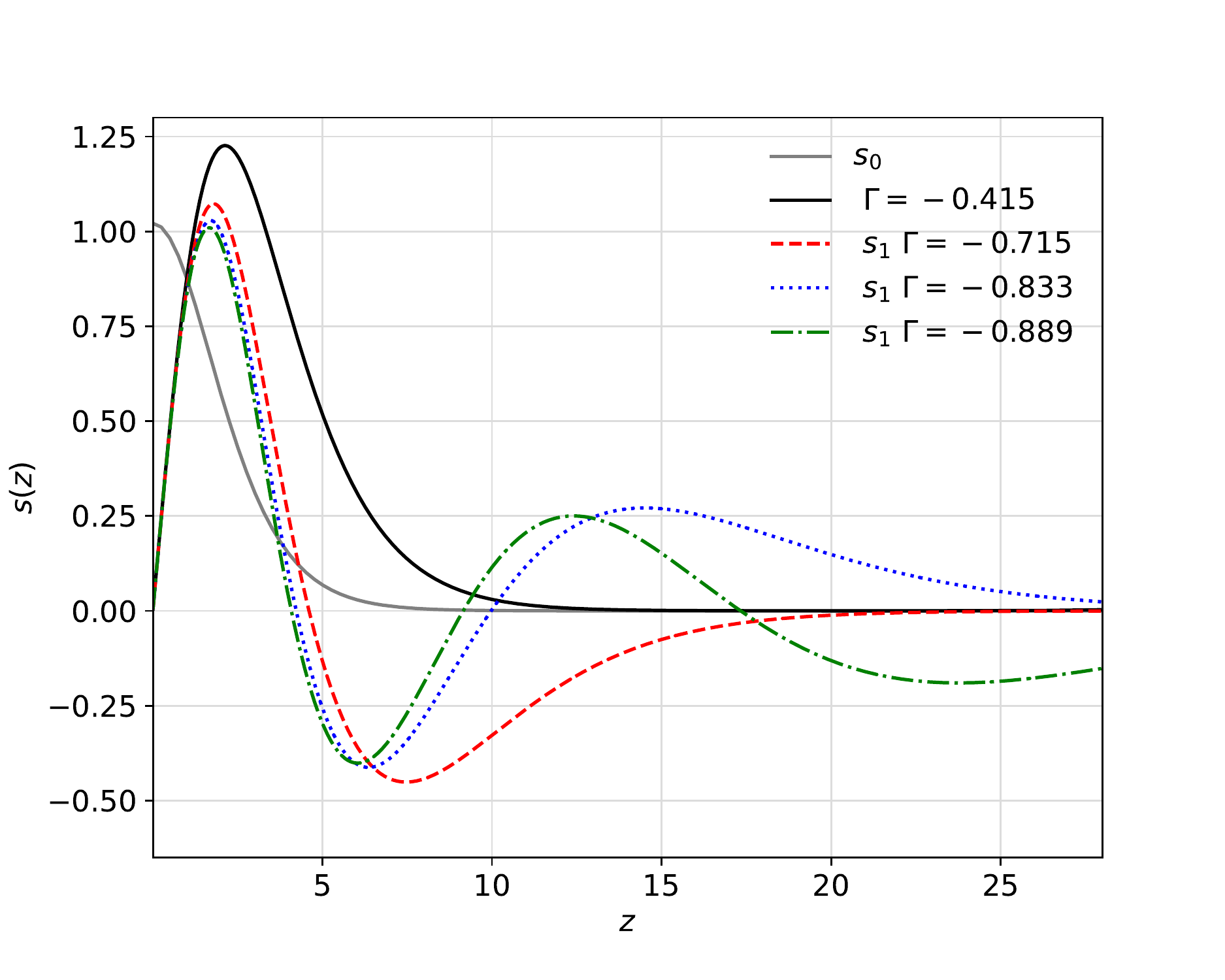}
\caption{Two-state solution. \textbf{Left:} The numerical solution (solid gray) and truncated series expansion of at small radius (dotted blue and dashed magenta) and large radii (dot-dashed red) as well as analytic Whittaker approximation (dotted black) for the wavefunction of a rotating perturbation with $\ell=1$ and $\Gamma=-0.415$ in the two-state ansatz.  
\textbf{Right:} Numerical solution for the wavefunction of a rotating perturbation with $\ell=1$ and $\Gamma=-0.415$ (black), $\Gamma  =-0.715$ (dashed red), $\Gamma=-0.833$ (dotted blue) and  $\Gamma=-0.889$ (dot-dashed green). The non-rotating ground state wavefunction is shown for comparisons (gray solid).
}
\label{fig:two_state}
\end{figure*}

As before, requiring the wavefunction to be normalizable implies that all coefficients $s^1_{0,m}$ vanish. The first non-vanishing terms appear for $n\!=\!1$, in which case we can simplify \cref{eq:oss_recs2} and write 
\be
	&2[\beta(1+\Gamma)^{-1/2}+\sigma'+m-2] s^1_{1,m-1} 
	\\
	& = [\ell(\ell+1) - (\sigma' + m - 2) (\sigma' + m - 3)] s^1_{1,m-2} \ .
	\label{eq:tss_recs_n2}
\ee
By setting $m = 1$, we obtain $\sigma'= 1 - \beta (1+\Gamma)^{-1/2}$ and note that $\sigma' \neq \sigma$ for $\Gamma \neq 0$, where $\sigma = 1-\beta$ appears in the non-rotating boson star expansion in \cref{largez-expansion}. Setting $m\!=\!M+1$ we find 
\be
    \!\!\!
    s^1_{1,M}  
	= \frac{[\ell(\ell\!+\!1) - (\sigma'\!+\!M\!-\!2)(\sigma'\!+\!M\!-\!1)] s^1_{1,M-1} }{2 M }. \!\!
\ee

Following the same procedure as in \cref{sec:oss_series}, we determine the expansion coefficient $s^1_{1,0}$ by matching the near field solution in \cref{smallz-expansion_oss2} truncated with $n \leq 100$ and $s_1^1=1$ and the far field solution in \cref{eq:rot_farexp2} truncated with $n \leq 1$, $m \leq 5$ at a matching point $z^*$ and obtain
\be
    s^1_{1,0} &= 4.910 \pm 0.074 \ . 
\label{eq:oss_matching2}
\ee
The uncertainty was estimated by performing multiple matchings for $3.2 \leq z^* \leq 3.5.$ 

\subsection{Leading Order Analytic Far-Field Solution}
\label{sec:oss_whittaker2}

Similar to the ground-state and the one-state solution, we can obtain an approximate analytical solution for the far field at leading order $n=1$. Using $v_0 \approx -1 + 2\beta z^{-1}$ and performing a change of variables to $w=2 z s_1$ and $y=2z (1+\Gamma)^{1/2}$, we can rewrite \cref{eq:structure_oss2} in the familiar Whittaker equation form 
\be
    \frac{d w^2}{dy^2} 
    + \left(-\frac{1}{4}+\frac{\kappa}{y} - \frac{\ell(\ell\!+\!1)}{y^2}\right) w 
    = 0
    \label{eq:inhom_whittaker2}
\ee
where $\kappa = \beta \cdot (1\!+\!\Gamma)^{-1/2}$. Note that \cref{eq:inhom_whittaker2} is homogeneous, while the Whittaker equation for the one-state ansatz in \cref{eq:inhom_whittaker} contained an additional inhomogeneous component (arising from the product $s_0 v_1$). The general solution to \cref{eq:inhom_whittaker2} is given by a linear combination of $W_{\kappa,\mu}(y)$ and $M_{\kappa,\mu}(y)$
\be
    w(y)=c \cdot W_{\kappa,\mu}(y) + c' \cdot M_{\kappa,\mu}(y)
\ee
where $\mu^2\!=\!\frac{1}{4} \!+\! \ell(\ell\!+\!1)$. 

The function $M_{\kappa,\mu}(y)$ diverges at large $y$ unless $\Gamma = \beta^2/\kappa^2-1$ with $\kappa$ being a natural number $\geq 2$. We will see in the next section that the solutions and corresponding values of $\Gamma$ do not fulfill this condition. Normalizability of the wavefunction then requires $c'\!=\!0$. The solution must therefore be solely described by $W_{\kappa,\mu}(y)$ which allows us to write
\be
    s^\text{far}_1 
    = \frac{c}{2z} W_{\kappa,\mu}(2z \sqrt{1+\Gamma}) \ . 
    \label{eq:oss_whittaker2}
\ee
Expanding this function leads to the ansatz in \cref{eq:rot_farexp2} and matching the coefficients of the leading terms allows us to identify $c=  s^1_{1,0} [2^{-\sigma'} \sqrt{1+\Gamma}]^{-1}$ with $ \sigma' = 1- \kappa$. 

\subsection{Numerical results}

As for the one state solution, we also obtain a numerical solution of \cref{eq:structure_oss2}, focusing on the case $\ell=1$. The equation is linear in $s_1$ which allows is to choose $s^1_1 = 1$ without loss of generality. We then use a Runge-Kutta~4 method to perform the numerical integration of \cref{eq:structure_oss2} and apply a shooting point method to find the values of $\Gamma$ for which the wavefunction converges at large radii. 

The lowest energy solution is obtained for $\Gamma=-0.415$, and the corresponding wavefunction is shown in the left panel of \cref{fig:one_state} as solid gray line. As before, we can fit the numerical solution with the far wavefunction given in \cref{eq:oss_whittaker2} and obtain
\be
    s^1_{1,0} &=4.894 \pm 0.013 \ . 
\ee
which agrees with our previous finding in \cref{eq:oss_matching2}. We also show the wavefunction profile using the truncated near solution in \cref{smallz-expansion_oss2} with $n \leq 15$ and $n \leq 100$, the far solution of \cref{eq:rot_farexp2} with $n\leq 1$ and  $m\leq 2$, and the Whittaker solution of \cref{eq:oss_whittaker2}. The truncated solution takes the form 
\be
    \hspace{-0.3cm}
    \!\!s_1 = 
    \begin{cases}
    z  - 0.135 z^{3} + 0.013 z^{5} & \\
    \ - 0.106 \!\cdot\! 10^{-2} z^{7} \,\; + 8.253 \!\cdot\! 10^{-5} z^{9} & \ \ \ \text{for}\\
    \ - 6.207 \!\cdot\! 10^{-6} z^{11} +  4.551 \!\cdot\! 10^{-7} z^{13} & z\!<\!3.45 \!\!\! \\
    \ - 3.275 \!\cdot\! 10^{-8} z^{15} & \\
    &\\
    (3.461 - 2.174 \,z^{-1} -1.153 \,z^{-2})   &
    \ \ \ \text{for}\\
    \ \times  \, z^{1.292} e^{- 0.765 z} &  z\!>\!3.45 \!\!\! \\
    \end{cases}
\ee
Again, with only a few terms in the expansion the wavefunction is described fairly well. 

The right panel of \cref{fig:two_state} shows additional numerical solutions for $\Gamma = -0.415$, $-0.715$, $-0.833$ and $-0.889$.  Note that these values of $\Gamma$ do not coincide with $\Gamma = \beta^2 / \kappa^2 - 1$ for $\kappa \geq 2$ like those of the one-state solutions. As argued before, this implies that far field solution is solely described by the Whittaker $W$ function.

\section{Conclusions}
\label{sec:conclusion}

Light scalar fields can form gravitationally bound compact objects, called boson stars. In the Newtonian limit, the profiles of boson stars are described by the Gross-Pitaevskii-Poisson equations.

In previous works, we presented a semi-analytic solution to these equations describing the profile of boson stars formed by  scalar fields \cite{Kling:2017mif,Kling:2017hjm}.  The solution was based on a series expansion which is parametrized by four expansion parameters that were obtained from numerical simulation at high accuracy. In this paper we have extended our methods to  find new solutions which allow for slowly rotating boson stars; specifically, we have found solutions for boson stars where the ratio of the angular momentum to the number of particles can be made arbitrarily small.

We considered two possibilities; in one  case, all the particles are in the same state and in the second  case the majority of the particles are in the zero angular momentum  ground state and a small number of particles are in an excited state containing angular momentum. In each case, we obtained accurate numerical and semi-analytic profiles (about $1 \%$ precision), thereby establishing the existence of these slowly rotating boson stars. 

The results and methods presented in this paper allow for systematic studies of the properties of boson stars in an analytic way without further relying on numerical simulations. There are several directions for further research; in particular, it would be interesting to extend these solutions to interacting scalars and to relativistic stars. It would also be interesting to see how the profiles are modified in the presence of other astrophysical objects like planets. We hope to return to these questions in future work.  
	
\bigskip 
\acknowledgments

This work of A.R. and F.R. was partially supported by the U.\,S.~National Science Foundation  under the award  NSF-PHY-1915005. F.K. is supported by U.\,S.~Department of Energy grant DE-AC02-76SF00515. F.R. was partially supported by the Division of Teaching Excellence and Innovation Graduate Fellowship at UCI. We are also grateful to the authors and maintainers of many open-source software packages from \texttt{Python}~\cite{van1995python}, including \texttt{numpy}~\cite{numpy, van2011numpy}, \texttt{matplotlib}~\cite{Hunter:2007}, \texttt{mpmath}~\cite{mpmath} and \texttt{scipy}~\cite{2020SciPy-NMeth} as well as \texttt{Jupyter} notebooks~\cite{soton403913}. 
	
	
\bibliography{main2}
	
\end{document}